\documentclass[conference,letterpaper]{IEEEtran}

\usepackage[utf8]{inputenc} 
\usepackage[T1]{fontenc}
\usepackage{url}
\usepackage{ifthen}
\usepackage[cmex10]{amsmath} 

\RequirePackage{amsfonts, amssymb, amsthm, amsopn}
\RequirePackage{amsmath}
\RequirePackage{bm} 
\RequirePackage{pifont}
\RequirePackage{mathrsfs}
\RequirePackage{fix-cm} 
\usepackage{xcolor}
\usepackage{xargs}
\usepackage[IEEEtran]{research18}
\usepackage{hyperref}
 \usepackage[caption=false,font=footnotesize]{subfig}

\newcommand{\Ma}{M}

\newcommand{\Rh}{R_\textnormal{h}}

\newcommandx{\strong}[3][2=\epsilon,3=n]{\mathcal{A}^{* (#3)}_{#2} (#1)}
\newcommandx{\weak}[3][2=\epsilon,3=n]{\mathcal{A}^{ (#3)}_{#2} (#1)}
\newcommandx{\typen}[2][2=n]{\top ^{ (#2)}_{#1}}
\newcommandx{\contypen}[3][3=n]{\top ^{ (#3)}_{#1}(#2)}

\newcommandx{\alltypen}[2][2=n]{\set P_{#2}({#1})}
\newcommandx{\allcontypen}[2][2=n]{\set P_{#2}({#1})}
\newcommandx{\typeseqn}[2][2=n]{\top ^{ (#2)}({#1})}
\newcommandx{\allprob}[1]{\set P({#1})}

\let\set\relax
\newcommand{\set}{\mathcal}

\newcommand{\bfyhat}{{\mathbf{\hat  y}}}
\newcommandx{\repbfyhat}[1][1=P]{\bfyhat_{#1}}
\newcommandx{\optrepbfyhat}[1][1=P]{\bfyhat^*_{#1}}

\newcommand{\bml}[1]{\mbox{\boldmath $ #1 $}}
\newcommand{\eqd}{\stackrel{\Delta}{=}}

\newcommand{\one}{\frac{1}{n}}
\newcommand{\half}{\frac{1}{2}}

\newcommand{\dist}{\mathsf D}

\newcommandx{\admchannel}[1][1=\dist]{\set W^{\leq #1} }
\newcommandx{\admchanneln}[1][1=n]{\set W^{\leq \dist}_{#1} }
\newcommandx{\discontypen}[2][2=\dist]{\set W^{\leq #2}_n ({#1})}

\newcommandx{\admchannelpermn}[1][1=n]{\bar {\set W}^{\leq \dist}_{#1} }

\begin{document}

\title{The State-Dependent Channel with a Rate-Limited Cribbing Helper}

\author{%
  \IEEEauthorblockN{Amos Lapidoth}
  \IEEEauthorblockA{Dept. of Information Technology and Electrical Engineering \\
                    ETH Zurich\\
                    8092 Zurich, Switzerland\\
                    Email: lapidoth@isi.ee.ethz.ch}
  \and
  \IEEEauthorblockN{Yossef Steinberg}
  \IEEEauthorblockA{Dept. of Electrical and Computer Engineering\\ 
                    Technion - IIT\\
                    Haifa 3200003, Israel\\
                    Email: ysteinbe@technion.ac.il}
}

\maketitle

\begin{abstract}
  The capacity of a memoryless state-dependent channel is derived for
  a setting in which the encoder is provided with rate-limited
  assistance from a cribbing helper that observes  the state sequence causally
  and the past channel inputs strictly-causally. Said cribbing may
  increase capacity but not to the level achievable by a
  message-cognizant helper.
\end{abstract}

\section{Introduction}\label{sec:intro}

An encoder for a state-dependent channel is said to have causal state
information if the channel input $X_{i}$ it produces at time~$i$ may
depend, not only on the message $m$ it wishes to transmit, but also on
the present and past channel states $S_{i}$ and $S^{i-1}$ (where
$S^{i-1}$ stands for the states $S_{1}, \ldots, S_{i-1}$). Its state
information is noncausal if, in addition to depending on the message,
the Time-$i$ input may depend on all the channel states: past
$S^{i-1}$, present $S_{i}$, and future $S_{i+1}^{n}$ (where $n$
denotes the blocklength, and $S_{i+1}^{n}$ stands for $S_{i+1},
\ldots, S_{n}$).

The former case was studied by Shannon \cite{Shannon:58p}, who showed
that capacity can be achieved by what-we-now-call Shannon
strategies. The latter was studied by Gel'fand and Pinsker
\cite{GelfandPinsker:80p3}, who showed that the capacity in this case
can be achieved using binning.

As of late, there has been renewed interest in the causal case, but
when the state information must be quantized before it is provided to
the encoder \cite{LapidothWang:23p}. While still causally, the encoder
isn't provided now with the state sequence $\{S_{i}\}$ directly, but
rather with some ``assistance sequence'' $\{T_{i}\}$ describing
it. Its Time-$i$ output $X_{i}$ is now determined by the message $m$
and by the present and past assistances $T^{i}$.
The assistance sequence is produced by a helper, which observes the
state sequence causally and produces the Time-$i$ assistance $T_{i}$
based on the present and past states $S^{i}$ subject to the additional
constraint that $T_{i}$ take values in a given finite set $\set{T}$
whose cardinality is presumably smaller than that of the state
alphabet $\set{S}$. (If the cardinality of $\set{T}$ is one, the
problem reduces to the case of no assistance; if it exceeds or equals
the cardinality of $\set{S}$, the problem reduces to Shannon's original
problem, because in this case $T_{i}$ can describe $S_{i}$
unambiguously.) We refer to the base-2 logarithm of the cardinality of
$\set{T}$ as the ``help rate'' and denote it $\Rh$:
\begin{IEEEeqnarray}{rCl}
  \Rh & = & \log_{2} |\set{T}|.
\end{IEEEeqnarray}

Three observations in \cite{LapidothWang:23p} inspired the present
paper:
\begin{enumerate}
\item Symbol-by-symbol quantizers are suboptimal: restricting $T_{i}$
  to be a function of $S_{i}$ may reduce capacity.
\item Allowing $T_{i}$ to depend not only on $S^{i}$ but also on the
  message $m$ may increase capacity.
\item If $T_{i}$ is allowed to depend on $S^{i}$ and the
  transmitted message, then message-cognizant symbol-by-symbol helpers
  achieve capacity: there is no loss in capacity in restricting
  $T_{i}$ to be a function of $(m,S_{i})$. 
\end{enumerate}

Sandwiched between the message-oblivious helper and the
message-cognizant helper is the cribbing helper: its Time-$i$
assistance $T_{i}$ depends on $S^{i}$ and on the past symbols produced
by the encoder
\begin{IEEEeqnarray}{rCl}
  T_{i} & = & T_{i}\bigl(S^{i},X^{i-1}\bigr).
\end{IEEEeqnarray}
Since one can reproduce the channel inputs from the states and
message, the cribbing helper cannot outperform the message-cognizant
helper. And since the helper can ignore the past channel inputs, the
cribbing capacity must be at least as high as that of the
message-oblivious helper.

The term ``cribbing'' is borrowed here from the seminal work of
Willems and van der Meulen in~\cite{WillemsVanderMeulen:85p} who
introduced it in the context of multiple-access channels. In this
context it was further studied in~\cite{AsnaniPermuter:13p} to account
for imperfect cribbing. 

Here we shall characterize the capacity with a cribbing helper and
show that the above inequalities can be strict: the message-cognizant
helper may outperform the cribbing helper, and the latter may
outperform the message-oblivious helper (presumably because, thanks
to the cribbing, it can learn something about the message). We further
show that the capacity of the cribbing helper can be achieved using a
Block-Markov coding scheme with backward decoding \cite{WillemsVanderMeulen:85p}.

It should be noted that message-cognizant helpers are advantageous
also in the noncausal case. For such helpers, capacity was recently
computed in \cite{LapidothWangYan:23a1} and
\cite{LapidothWangYan:23a2}. Cribbing, however, is somewhat less natural
in this setting.

\section{Problem Statement and Main Result}
\label{sec:main_result}
We are given a state-dependent discrete memoryless channel
$W_{Y|XS}$ 
of finite input, output, and state alphabets $\set{X}$,
$\set{Y}$, and $\set{S}$. When its input is $x \in \set{X}$ and its state
is $s \in \set{S}$, the probability of its output being $y \in \set{Y}$
is $W_{Y|XS}(y|x,s)$.
Its states $\{S_{i}\}$ are drawn IID $\sim P_{S}$, where $P_{S}$ is
some given probability mass function (PMF) on the state alphabet
$\set{S}$. Also given is some finite set $\set{T}$ we call the
description alphabet. We shall assume throughout that its cardinality
is at least two
\begin{IEEEeqnarray}{rCl}
  |\set{T}| & \geq & 2
\end{IEEEeqnarray}
because otherwise the helper cannot provide any assistance.

Given some blocklength $n$, a rate-$R$ message set is a set $\set{M}$
whose cardinality is $2^{nR}$ (where we ignore the fact that the
latter need not be an integer).

A blocklength-$n$ encoder for our channel comprises $n$ mappings
\begin{IEEEeqnarray}{rCl}
  f_{i}  \colon \set{M} \times \set{T}^{i} \to \set{X}, \qquad i = 1,
  \ldots, n \label{eq:def_encoder_intro1}
\end{IEEEeqnarray}
with the understanding that if the message to be transmitted is $m \in \set{M}$,
and if the assistance sequence produced by the helper is $t^{n} \in
\set{T}^{n}$, then the Time-$i$ channel input produced by the encoder
is
\begin{IEEEeqnarray}{rCl}
  x_{i} & = & f_{i}(m,t^{i}) \label{eq:def_encoder_intro2}
\end{IEEEeqnarray}
which we also denote $x_{i}(m,t^{i})$. Here $\set{T}^{i}$ denotes the
$i$-fold Cartesian product
\begin{IEEEeqnarray}{rCl}
  \set{T}^{i} & = & \underbrace{\set{T} \times \set{T} \times \cdots
    \times \set{T}}_{\text{$i$ times}}
\end{IEEEeqnarray}
and $t^{j}$ denotes $t_{1}, \ldots, t_{j}$.

A blocklength-$n$ cribbing helper comprises $n$ mapping
\begin{IEEEeqnarray}{rCl}
  h_{i}\colon \set{X}^{i-1} \times \set{S}^{i} \to \set{T}, \qquad i =
  1, \ldots, n
\end{IEEEeqnarray}
with the understanding that---after observing the channel inputs
$x_{1}, \ldots, x_{i-1}$ and the states $s_{1}, \ldots, s_{i}$---the
helper produces the Time-$i$ assistance
\begin{IEEEeqnarray}{rCl}
  t_{i} & = & h_{i}\bigl(x^{i-1},s^{i}\bigr)\label{eq:ti_definition}
\end{IEEEeqnarray}
which we also denote $t_{i}\bigl(x^{i-1},s^{i}\bigr)$.

Communication proceeds as follows: the helper produces the
Time-$1$ assistance $t_{1}$ that is given by $h_{1}(s_{1})$, and the
encoder then produces the first channel input
$x_{1} = f_{1}(m,t_{1})$. The helper then produces the Time-$2$
assistance $t_{2}$ that is given by $h_{2}(x_{1}, s^{2})$, and the
encoder then produces the second channel input
$x_{2} = f_{2}(m,t^{2})$, and so on.

The decoder is cognizant neither of the state sequence $s^{n}$ nor of
the assistance sequence $t^{n}$: it is thus a mapping of the form
\begin{equation}
  \label{eq:def_decoder}
  \phi \colon \set{Y}^{n} \to \set{M}
\end{equation}
with the understanding that, upon observing the output sequence
$Y^{n}$, the decoder guesses that the transmitted message is
$\phi(Y^{n})$

Let $P_\textnormal{e}$ denote the probability of decoding error,
averaged over all the messages. If $P_\textnormal{e} < \eps$, then we
say that the coding scheme is of parameters
$(n,2^{nR},|\set{T}|,\epsilon)$ or that it is a
$(n,2^{nR},|\set{T}|,\epsilon)$-scheme.

A rate $R$ is said to be achievable if for every $\epsilon>0$ there
exist, for all sufficiently large $n$, schemes as above with
$P_{\textnormal{e}} < \eps$.
The capacity of the channel is defined as the supremum of all
achievable rates $R$, and is denoted $C$.

Define
\begin{equation}
\label{eq:def_ci}
C^{(I)} = \max \min \bigl\{ I(UV;Y), I(U;X|VT) \bigr\}
\end{equation}
where the maximum is over all finite sets $\set{U}$ and $\set{V}$ and
over all joint distributions of the form
\begin{equation}
\label{eq:def_ci_joint}
P_S \, P_{UV} \, P_{T|VS} \, P_{X|UVT} \, W_{Y|XS}
\end{equation}
with $T$ taking values in the assistance alphabet ${\cal T}$. 
Our main result is stated next:
\begin{theorem}
\label{theo:main}
The capacity $C$ of the memoryless state dependent channel with a rate-limited
 cribbing helper  equals $C^{(I)}$:
\begin{equation}
\label{eq:main}
C=C^{(I)}.
\end{equation}
Moreover, the maximum in~\eqref{eq:def_ci} can be achieved when:
\begin{enumerate}
\item $P_{T|VS}$ and $P_{X|UVT}$ are $0-1$ laws;
\item The alphabet sizes of $U$ and $V$ are restricted to
\begin{IEEEeqnarray}{rCl}
|{\cal V}| &\leq& L^2 \, |{\cal S}| \, \bigl( |{\cal T}|-1 \bigr) + L\nonumber\\
|{\cal U}| &\leq& L^3 \, |{\cal T}| \bigl( |{\cal X}|-1 \bigr) +L\nonumber
\end{IEEEeqnarray}   
where $L= |{\cal X}| \, |{\cal T}| \, |{\cal S}| + 1$
\item The chain $V\markov U\markov (XTS)\markov Y$ is a Markov chain.
\label{item:markov_theo_main}
\end{enumerate}
\end{theorem}
The proof is given in Section~\ref{sec:proof_main}.

\section{Example}
We next present an example where the message-cognizant helper
outperforms the cribbing helper, and the latter outperforms the plain
vanilla causal helper. It is trivial to find cases where the three
perform identically, e.g., when the state does not affect the channel.
The example is borrowed from \cite[Example~7]{LapidothWang:23p} (from
which we also lift the notation).

The channel inputs, states, and outputs are binary tuples
\begin{equation}
  \set{X} = \set{S} = \set{Y} = \{0,1\} \times \{0,1\}
\end{equation}
and are denoted $(A,B)$, $\bigl( S^{(0)}, S^{(1)}\bigr)$, and
$\bigl( Y^{(0)}, Y^{(1)} \bigr)$. The two components of the state are
IID, each taking on the values $0$ and $1$ equiprobably. Given the
state and input, the channel output is deterministcally given by
  \begin{equation}
    Y = \bigl( A, B \oplus S^{(A)} \bigr).
  \end{equation}
The assistance is one-bit assistance, so $\set{T} = \{0,1\}$. 

As shown in \cite[Claim~8]{LapidothWang:23p}, the capacity with a
message-cognizant helper is $2$ bits, and with a message-oblivious
helper $\log 3$. Here we show that the capacity with a cribbing helper
is strictly smaller than $2$ bits and strictly larger than $\log
3$. All logarithms in this section are base-2 logarithms, and all
rates are in bits.

We begin by showing the former. Recall the constraints
\begin{IEEEeqnarray}{rCl}
  R & \leq & I(UV;Y) \label{eq:cons1} \\
  R & \leq & I(U;X|VT) \label{eq:cons2}
\end{IEEEeqnarray}
the form of the joint PMF
\begin{equation}
  \label{eq:PMFform}
  P_{S} \, P_{V} \, P_{T|VS} \, P_{U|V} P_{X|UVT} \, W_{Y|XS}
\end{equation}
and that we may assume that $P_{X|UVT}(x|u,v,t)$ is a $0-1$ valued.

Note that~\eqref{eq:PMFform} implies
\begin{equation}
  \label{eq:amos_Markov4}
  ST \markov V \markov U
\end{equation}
and consequently
\begin{equation}
  \label{eq:amos_Markov4prime}
  S \markov TV \markov U.
\end{equation}

We will show that the above constraints cannot be both satisfied if $R
= 2$. To that end, we assume that 
\begin{IEEEeqnarray}{rCl}
  \label{eq:consRHS2}
  I(U;X|VT) & = & 2 
\end{IEEEeqnarray}
(it cannot be larget because $\card{\set{X}} = 4$) and prove that
\begin{IEEEeqnarray}{rCl}
  \label{eq:consRHS1}
  I(UV;Y) & < & 2. 
\end{IEEEeqnarray}
Since $\set{Y}$ is of cardinality $4$, it suffices to show that
\begin{equation}
  \label{eq:amos_condYgUV1}
  H(Y|UV) > 0.
\end{equation}
In fact, it suffices to show that
\begin{equation}
  \label{eq:amos_condYgUVT2}
  H(Y|UVT) > 0,
\end{equation}
i.e., that there exist $u^{\star},v^{\star},t^{\star}$ of positive
probability for which
\begin{equation}
  \label{eq:amos_condYgUVT2}
  H(Y|U=u^{\star}, V = v^{\star}, T = t^{\star}) > 0.
\end{equation}

Since $\card{\set{X}} = 4$, \eqref{eq:consRHS2} implies that
\begin{equation}
  \label{eq:condXuniform}
  P_{X|V=v, T=t} \quad \textnormal{is uniform $\forall (v,t)$}.
\end{equation}

Fix any $v^{\star}$ (of positive probability). As we next argue, there
must exist some $t^{\star}$ for which $P_{S|V = v^{\star}, T =
  t^{\star}}$ is not zero-one valued. Indeed, by~\eqref{eq:PMFform},
$V \indep S$, so $H(S|V = v^{\star}) = H(S) = 2$ and
\begin{IEEEeqnarray}{rCl}
  H(S|T,V = v^{\star}) & = & H(S|V = v^{\star}) - I(S;T|V =
  v^{\star})\\
  & = & H(S) - I(S;T|V = v^{\star})\\
  & \geq & 2 - \log \card{\set{T}} \\
  & = & 1 
\end{IEEEeqnarray}
so there must exist some $t^{\star}$ for which
\begin{equation}
  \label{eq:HSlarger1}
  H(S|V = v^{\star}, T = t^{\star}) \geq 1.
\end{equation}

Conditional on $V = v^{\star}, T = t^{\star}$, the chance variable $U$
has some PMF $P_{U|V = v^{\star}, T = t^{\star}}$ (equal to $P_{U|V =
  v^{\star}}$ by~\eqref{eq:PMFform}) under which
$X(U,v^{\star},t^{\star})$ is uniform;
 see~\eqref{eq:condXuniform}. It follows that there exist $u_{0}$ and
$u_{1}$ (both of positive conditional probability) such that
\begin{IEEEeqnarray}{rCl}
  A(u_{0},v^{\star},t^{\star}) & = & 0 \\
  A(u_{1},v^{\star},t^{\star}) & = & 1 
\end{IEEEeqnarray}
where we introduced the notation
\begin{equation}
  X(u,v^{\star},t^{\star}) = \bigl( A(u,v^{\star},t^{\star}), B(u,v^{\star},t^{\star})\bigr).
\end{equation}

Returning to~\eqref{eq:HSlarger1}, we note that it implies that
\begin{IEEEeqnarray}{rCl}
 H\bigl( S^{(0)} \big| V = v^{\star}, T = t^{\star} \bigr) > 0
 \end{IEEEeqnarray}
 or
 \begin{IEEEeqnarray}{rCl}
  H\bigl( S^{(1)} \big| V = v^{\star}, T = t^{\star} \bigr) > 0.
\end{IEEEeqnarray}
In the former case $H(Y|U=u_{0}, V = v^{\star}, T = t^{\star})$ is
positive, and in the latter
$H(Y|U=u_{1}, V = v^{\star}, T = t^{\star})$ is positive. This
establishes the existence of a triple $(u^{\star},v^{\star},
t^{\star})$ for which~\eqref{eq:amos_condYgUVT2} holds, and thus
concludes the proof.

We next show that the capacity with a cribbing helper exceeds $\log 3$. 
Let
\begin{equation}
 U=(A,\tilde{U}) 
\end{equation}
be uniform over $\{0,1\}\times\{0,1\}$, and let $\sigma$ be a
Bernoulli-$\alpha$ 
random variable that is independent of $U$ and of the channel,
for some $\alpha \in [0,1]$ to be specified later.

Define the random variables
\begin{IEEEeqnarray}{rCl}
\label{eq:example_Vtilde_def}
\tilde{V} &=& \left\{ \begin{array}{lcl}
                             A & \mbox{if} & \sigma=1\\
                             0 \  & \mbox{if} & \sigma=0\\
                             \end{array} \right.
\end{IEEEeqnarray}
and
\begin{equation}
\label{eq:example_V_def}
V=(\tilde{V},\sigma).
\end{equation}
Let $h(s,v)$---which can also be written as  $h\bigl( (s^{(0)},s^{(1)}), (\tilde{v}, \sigma)
\bigr)$---equal $s^{(\tilde{v})}$, i.e., 
\begin{equation}
  \label{eq:example_T_def}
  T = S^{(\tilde{V})}
\end{equation}
so
\begin{equation}
\label{eq:example_T_def2}
T= \left\{ \begin{array}{ll}
  S^{(A)} & \mbox{w.p. $\alpha$}\\
   S^{(0)}  & \mbox{w.p. $1-\alpha$.}\\
        \end{array} \right. 
\end{equation}
Let the encoder function $f(u,v,t)$ ignore $v$ and  result
in 
\begin{equation}
\label{eq:example_X_def}
X^{(0)} = A,\quad X^{(1)} = \tilde{U}\oplus T
\end{equation}
where $X=(X^{(0)},X^{(1)})$. More explicitly,
\begin{equation}
  f\bigl( (A,\tilde{U}), T \bigr) = \bigl( A, \tilde{U}\oplus T \bigr).
\end{equation}
Note that with the variables defined in~(\ref{eq:example_Vtilde_def})-(\ref{eq:example_X_def}),
the Markov relations in item~\ref{item:markov_theo_main} of Theorem~\ref{theo:main} hold.

We proceed to calculate the rate bounds. For~(\ref{eq:cons1}) we have
\begin{IEEEeqnarray}{rCl}
I(UV;Y) &=& I(U\tilde{V}\sigma;Y)\geq I(U\tilde{V};Y|\sigma)\nonumber\\
             &=& \alpha \, I(U\tilde{V};Y|\sigma=1) + (1-\alpha) \, I(U\tilde{V};Y|\sigma=0)\nonumber\\
             &=& \alpha \, I(A\tilde{U};Y|\sigma=1)\nonumber\\
             & &\mbox{} +(1-\alpha)   \, I(A\tilde{U};Y|\sigma=0).\label{eq:example_I_UV_Y}
\end{IEEEeqnarray}
We next evaluate each of the terms on RHS separately. When $\sigma=1$,
\begin{IEEEeqnarray}{rCl}
\label{eq:example_channel_sigma1}
T &=& S^{(A)}\nonumber\\
X^{(1)} &=& \tilde{U}\oplus S^{(A)}\nonumber\\
Y^{(1)}  &=& X^{(1)}\oplus S^{(A)} = \tilde{U}\oplus S^{(A)}\oplus S^{(A)} = \tilde{U}
\end{IEEEeqnarray}
hence
\begin{equation}
Y = (Y^{(0)},Y^{(1)})= (A,\tilde{U})\label{eq:example_AtildeU}
\end{equation}
implying
\begin{equation}
\label{eq:example_I_UV_Y_1}
I(A\tilde{U};Y|\sigma=1) = H(Y) = 2. 
\end{equation}

When $\sigma=0$, 
\begin{IEEEeqnarray}{rCl}
T&=& S^{(0)}\nonumber\\
X &=& (A, \tilde{U}\oplus S^{(0)} )\nonumber\\
Y &=& (A, \tilde{U}\oplus S^{(0)}\oplus S^{(1)} ) \label{eq:example_channel_sigma0}
\end{IEEEeqnarray}
so 
\begin{IEEEeqnarray}{rCl}
I(A\tilde{U};Y|\sigma=0) &=& I(A\tilde{U};Y^{(0)}Y^{(1)}|\sigma=0)\nonumber\\
    & =& I(A\tilde{U};A,\tilde{U}\oplus S^{(0)}\oplus S^{(A)})
\nonumber\\
 &=& I(A\tilde{U};A) \nonumber\\
   & &\mbox{} + I(A\tilde{U};A,\tilde{U}\oplus S^{(0)}\oplus S^{(A)}|A)\nonumber\\
 &=& H(A) +\half \, I(\tilde{U};\tilde{U}\oplus S^{(0)}\oplus S^{(0)}
 |A=0)\nonumber\\
 & &\mbox{} +\half \, I(\tilde{U};\tilde{U}\oplus S^{(0)}\oplus S^{(1)} |A=1)
 \nonumber\\
&= &  H(A) + \half H(\tilde{U}) +0 = \frac{3}{2}.
 \label{eq:example_I_UV_Y_0}
\end{IEEEeqnarray}
From~(\ref{eq:example_I_UV_Y_1}), (\ref{eq:example_I_UV_Y_0}),
and~(\ref{eq:example_I_UV_Y}) we obtain that the RHS of~(\ref{eq:cons1}) satisfies
\begin{equation}
I(UV;Y) \geq 2\alpha +(1-\alpha)\frac{3}{2} = (\alpha + 3)/2.
\label{eq:example_I_UV_Y_final}
\end{equation}
Next we evaluate the RHS of~(\ref{eq:cons2}):
\begin{IEEEeqnarray}{rCl}
I(U;X|VT) &=& I(U;X|\tilde{V},\sigma,T)\nonumber\\
                &=& \alpha \, I(U;X|\tilde{V},\sigma=1,T)\nonumber\\
                & &\mbox{} + (1-\alpha)
                \, I(U;X|\tilde{V},\sigma=0,T)\nonumber\\
                &=& \alpha \, I(A\tilde{U};X|A,S^{(A)},\sigma=1)\nonumber\\
                & &\mbox{} +
                (1-\alpha) \, I(A\tilde{U};A,\tilde{U}\oplus A^{(0)}|S^{(0)},\sigma=0)
\nonumber\\
                &=& \alpha \, I(\tilde{U}; A,\tilde{U}\oplus T|A S^{(A)},\sigma=1)\nonumber\\
                & &\mbox{} + 
                         (1-\alpha) \, I(A\tilde{U};A,\tilde{U}\oplus S^{(0)} | S^{(0)},\sigma=0)\nonumber\\
                 &=&    \alpha \, I(\tilde{U}; \tilde{U}\oplus T|A
                 S^{(A)},\sigma=1)\nonumber\\
                 & &\mbox{}  + (1-\alpha) \, H(A,\tilde{U})\nonumber\\
                 &=& \alpha \, H(\tilde{U}) + H(A,\tilde{U})\nonumber\\
                 & =& \alpha + (1-\alpha)2 = 2-\alpha.
                 \label{eq:example_I_U_X_1}       
\end{IEEEeqnarray}
In view of~(\ref{eq:example_I_UV_Y_final}) and~(\ref{eq:example_I_U_X_1}), any rate $R$ satisfying
\begin{equation}
R \leq \min\{(\alpha + 3)/2, 2-\alpha\} \label{eq:example_finalAlpha}
\end{equation}
is achievable. Choosing $\alpha = 1/3$ (which maxmizes the RHS
of~(\ref{eq:example_finalAlpha})),
demonstrates the achievability of
\begin{equation}
  R =  5/3 
  \label{eq:example_finalAchievable}
\end{equation}
which exceeds $\log 3$.

\section{Proof of Theorem~\ref{theo:main}}
\label{sec:proof_main}
\subsection{Direct Part}
\label{subsec:proof_main_direct}

Pick a distribution as in~\eqref{eq:def_ci_joint}, where 
$P_{T|SV}$ and $P_{X|UVT}$ are 0-1 laws, so
\begin{IEEEeqnarray}{rCl}
x &=& f(u,v,t)\label{eq:def_f}\\
t &=& h(s,v) \label{eq:def_h}
\end{IEEEeqnarray}
for some deterministic functions $f$ and $h$.
Extend these functions to act on $n$-tuples components-wise so that if
$\bml{s}, \bml{v}$ are $n$-tuples in $\set{S}^{n}$ and $\set{V}^{n}$,
then $\bml{t} = h(\bml{s},\bml{v})$ indicates that $\bml{t}$ is an
$n$-tuple in $\set{T}^{n}$ whose $i$-th component $t_{i}$ is
$h(s_{i},v_{i})$, where $s_{i}$ and $v_{i}$ are the corresponding
components of~$\bml{s}$ and~$\bml{v}$. Likewise we write
$\bml{x} = f(\bml{u},\bml{v},\bml{t})$.

To prove achievability, we propose a block-Markov coding scheme with
the receiver performing backward decoding. Although only the receiver
is required to decode the message, in our scheme the helper does so too
(but not with backward decoding, which would violate causality).

The transmission comprises $B$ $n$-length sub-blocks, for a total of $Bn$
channel uses. The transmitted message $m$ is represented by
$B-1$ sub-messages $m_{1}, \ldots, m_{B-1}$, with each of the
sub-messages taking values in the set
${\cal M}\eqd\{1,2,\ldots,2^{nR}\}$. The overall transmission rate is
thus $R(B-1)/B$, which can be made arbitrarily close to $R$ by
choosing $B$ very large. The $B-1$ sub-messages are transmitted in the
first $B-1$ sub-blocks, with $m_{b}$ transmitted in Sub-block~$b$ (for
$b \in [1:B-1]$). Hereafter, we use
$\bml{s}^{(b)}$ to denote the state $n$-tuple affecting the channel in
Sub-block~$b$ and use $s_i^{(b)}$ to denote its $i$-component (with
$i \in [1:n]$).  Similar notation holds for $\bml{x}^{(b)}$,
$\bml{y}^{(b)}$, etc.

We begin with an overview of the scheme, where we focus on the
transmission in Sub-blocks $2$ through $B-1$: the first and last
sub-blocks must account for some edge effects that we shall discuss
later. Let $b$ be in this range. The coding we use in Sub-block~$b$ is
superposition coding with the cloud center determined by $m_{b-1}$ and
the satellite by $m_{b}$.

Unlike the receiver, the helper, which must be causal, cannot employ
backward decoding: it therefore decodes each sub-message at the end of the
sub-block in which it is transmitted. Consequently, when Sub-block~$b$
begins, it already has a very reliable guess $\hat{m}_{b-1}$ of
$m_{b-1}$ (based on the previous channel inputs $\bml{x}^{(b-1)}$ it
cribbed). The encoder, of course, knows $m_{b-1}$, so the two can
agree on the cloud center $\bml{v}^{(b)}(m_{b-1})$ indexed by
$m_{b-1}$. (We ignore for now the fact that $\hat{m}_{b-1}$ may, with
small probability, differ from $m_{b-1}$.) The satellite is computed
by the encoder as $\bml{u}^{(b)}(m_{b}|m_{b-1})$; it is unknown to the
helper. The helper produces the Sub-block~$b$ assistance
$\bml{t}^{(b)}$ based on the state sequence and the cloud center
\begin{equation}
\bml{t}^{(b)} = h\bigl(\bml{s}^{(b)}, \bml{v}^{(b)}(m_{b-1})\bigr).  
\end{equation}
(Since $h(\cdot,\cdot)$ acts componentwise, this help is causal with
the $i$-th component of $\bml{t}^{(b)}$ being a function of the
corresponding component $s^{(b)}_{i}$ of the state sequence and
$\bml{v}^{(b)}(m_{b-1})$; it does not require knowledge of future
states.)

For its part, the encoder produces the $n$-tuple 
\begin{equation}
   \bml{x}^{(b)} = f\bigl( \bml{u}^{(b)}(m_{b}|m_{b-1}), \bml{v}^{(b)}(m_{b-1}), \bml{t}^{(b)}\bigr)
\end{equation}
with causality preserved because
$\bml{u}^{(b)}(m_{b}|m_{b-1})$ and $\bml{v}^{(b)}(m_{b-1})$ can be computed
from $m_{b-1}$ and $m_{b}$ ahead of time, and because $\bml{t}$ is
presented to the encoder causally and $f(\cdot)$ operates
component-wise.

As to the first and last sub-blocks: In the first we set $m_{0}$ as
constant (e.g., $m_{0} = 1$), so we have only one cloud center. And in
Sub-block~$B$ we send no fresh information, so each cloud center has
only one sattelite.

We now proceed to a more formal exposition. For this, we will need
some notation. Given a joint distribution $P_{XYZ}$, we denote by
${\cal T}_{XY}$ the set of all jointly typical sequences
$(\bml{x},\bml{y})$ where the length $n$ is understood from the
context, and we adopt the $\delta$-convention
of~\cite{CsiszarKorner:82b}.  Similarly, given a sequence $\bml{z}$,
${\cal T}_{XYZ}(\bml{z})$ stands for the set of all pairs
$(\bml{x},\bml{y})$ that are jointly typical with the given squence
$\bml{z}$.

To describe the first and last sub-blocks, we define $m_0=1$, and
$m_B=1$.  The proof of the direct part is based on random
coding and joint typicality decoding.

\subsubsection{Code Construction}
\label{subsubsec:code_construction}
We construct $B$ codebooks ${\cal C}_b$, $b\in[1:B]$, each of length $n$.
Each codebook ${\cal C}_b$, $b\in[1:B]$, is generated randomly and  independently of the other codebooks,
as follows:
\begin{itemize}
\item For every $b\in[1:B]$, generate $2^{nR}$ length $n$ cloud centers $\bml{v}^{(b)}(j)$, $j\in{\cal M}$,
independently of each other, and iid according  to $P_V$.
\item For every  $b\in[1:B]$ and $j\in{\cal M}$, generate $2^{nR}$
  length-$n$ satellites 
$\bml{u}^{(b)}(m|j)$, $m\in {\cal M}$,  independently, each according to 
\begin{equation}
\prod_{i=1}^n P_{U|V}(\cdot| v_i^{(b)}(j)). \label{eq:u_codewords_def}
\end{equation}
\end{itemize}
The codebook ${\cal C}_b$ is the collection
\begin{equation}
\label{eq:Cb}
\left\{ \bml{v}^{(b)}(j), \bml{u}^{(b)}(m|j), j\in{\cal M}, m\in{\cal M}\right\}
\end{equation}
Reveal the codebooks to the encoder, decoder, and helper. 

\subsubsection{Operation of the code}
\label{subsubsec:code_operation}
We first describe the operation of the helper and encoder at the first
sub-block.

\noindent
\underline{Helper}. At the first sub-block, $b=1$, the helper produces
\begin{equation}
\label{eq:code_helper_1}
\bml{t}^{(1)} = (t_1^{(1)},t_2^{(1)},\ldots,t_n^{(1)})
\end{equation}
where
\begin{equation}
\label{eq:code_helper_2}
t_i^{(1)} = h(s_i^{(1)},v_i^{(1)}(m_0)), \quad 1\leq i\leq n.
\end{equation}
Note that $\bml{t}^{(1)}$ is causal in $\bml{s}^{(1)}$.

\noindent
\underline{Encoder}. Set $\bml{u}^{(1)} = \bml{u}^{(1)}(m_1|m_0)$ and $\bml{v}^{(1)} = \bml{v}^{(1)}(m_0)$.
The input to the channel is
\begin{equation}
\label{eq:code_encoder_1}
\bml{x}^{(1)} = \left(x_1^{(1)},x_2^{(1)},\ldots,x_n^{(1)}\right)
\end{equation}
where
\begin{IEEEeqnarray}{rCl}
\label{eq:code_encoder_2}
x_i^{(1)} &=& f\left(u_i^{(1)}(m_1|m_0),v_i^{(1)}(m_0),t_i^{(1)}\left(s_i^{(1)},v_i^{(1)}(m_0)\right)\right)\nonumber\\
             &=& f\left(u_i^{(1)},v_i^{(1)},t_i^{(1)}\right), \quad 1\leq i\leq n.
\end{IEEEeqnarray}
Note that $\bml{x}^{(1)}$ is causal in $\bml{t}^{(1)}$.

\noindent
\underline{Helper at the end of the sub-block}.
Thanks to its cribbing, at the end of Sub-block~$1$ the helper is
cognizant of $\bml{x}^{(1)}$. 
In addition, it knows  $\bml{v}^{(1)}$ (since it is determined by $m_0$, which was set a-priori) 
and $\bml{t}^{(1)}$ (since it was produced by itself). 
The helper now decodes the message $m_1$ by looking for an index $j\in{\cal M}$
such that 
\begin{equation}
\label{eq:decoding_at_helper}
\left(\bml{u}^{(1)}(j|m_0),\bml{x}^{(1)}\right) \in T_{UXVT}(\bml{v}^{(1)},\bml{t}^{(1)}).
\end{equation}
If such an index $j$ exists and is unique, the helper sets
$\hat{m}_1=j$. Otherwise, an error is declared.  By standard results,
the probability of error is vanishingly small provided that
\begin{equation}
\label{eq:code_helper_end1}
R < I(U;X|VT).
\end{equation}
Denote by $\hat{m}_1$ the message decoded by the helper at the end of Sub-block 1.
We proceed to describe the operation of the helper and encoder in Sub-block $b$, $2\leq b\leq B-1$.

\noindent
\underline{Helper, $2\leq b\leq B-1$}. Denote by $\hat{m}_{b-1}$ the message decoded by the helper
at the end of Sub-block $b-1$. In Sub-block $b$, the helper produces
\begin{equation}
\label{eq:code_helper_1_b}
\bml{t}^{(b)} = (t_1^{(b)},t_2^{(b)},\ldots,t_n^{(b)})
\end{equation}
where
\begin{equation}
\label{eq:lcode_helper_2_b}
t_i^{(b)} = h(s_i^{(b)},v_i^{(b)}(\hat{m}_{b-1})), \quad 1\leq i\leq n.
\end{equation}

\noindent
\underline{Encoder, $2\leq b\leq B-1$}. 
Set $\bml{u}^{(b)} = \bml{u}^{(b)}(m_b|m_{b-1})$ and $\bml{v}^{(b)} = \bml{v}^{(b)}(\hat{m}_{b-1})$.
The input to the channel is
\begin{equation}
\label{eq:code_encoder_1_b}
\bml{x}^{(b)} = \left(x_1^{(b)},x_2^{(b)},\ldots,x_n^{(b)}\right)
\end{equation}
where
\begin{IEEEeqnarray}{rCl}
\label{eq:code_encoder_2_b}
x_i^{(b)} &=& f\left(u_i^{(b)}(m_b|m_{b-1}),v_i^{(b)}(m_{b-1}),t_i^{(b)}\left(s_i^{(b)},v_i^{(b)}(\hat{m}_{b-1})\right)\right)
                          \nonumber\\
             &=& f\left(u_i^{(b)},v_i^{(b)},t_i^{(b)}\right), \quad 1\leq i\leq n.
\end{IEEEeqnarray}
Note that $\bml{t}^{(b)}$ and $\bml{x}^{(b)}$ are causal in $\bml{s}^{(b)}$ and $\bml{t}^{(b)}$, respectively.  

\noindent
\underline{Helper at the end of the sub-block, $2\leq b\leq B-1$}.  At
the end of Sub-block $b$ the helper has $\bml{x}^{(b)}$ at hand.  In
addition, it has $\bml{v}^{(b)}(\hat{m}_{b-1})$ (since $\hat{m}_{b-1}$
was decoded at the end of the previous sub-block) and $\bml{t}^{(b)}$
(since it was produced by itself).  The helper now decodes the message
$m_b$.  Assuming that $\hat{m}_{b-1}$ was decoded correctly, this can
be done with low probability of error if~(\ref{eq:code_helper_end1})
is satisfied.

We proceed to the last sub-block, where no fresh information is sent. Here $m_B=1$, and  the operations of the
helper and encoder proceed exactly as in~(\ref{eq:code_helper_1_b})--(\ref{eq:code_encoder_2_b}), with $b=B$.
Note that in Sub-block $B$ the helper need not decode $m_B$ since it is set a-priori and known to all. 

\subsubsection{Decoding}
\label{subsubsec:decoding}
At the destination we employ backward decoding. Starting at Sub-block $B$ with $m_B=1$, the encoder looks for 
an index $j\in{\cal M}$ such that
\begin{equation}
\label{eq:decoding_1}
(\bml{u}^{(B)}(1|j),\bml{v}^{(B)}(j),\bml{y}^{(B)}) \in {\cal T}_{UVY}
\end{equation}
If such index exists and is unique, 
the decoder sets $\hat{\hat{m}}_{B-1}=j$. Otherwise, an error is declared. By standard result, the decoding is correct
with probability approaching~1 provided
\begin{equation}
\label{eq:decoding_rate_bound}
R<I(UV;Y).
\end{equation}
For blocks $B-1$, $B-2\ldots$, the decoding proceeds as in~(\ref{eq:decoding_1}), 
with the exception that the
{\it estimate} $\hat{\hat{m}}_b$ replaces the default value $m_B=1$ in~(\ref{eq:decoding_1}).
Thus, in Sub-block $B-1$, the decoder has at hand the estimate $\hat{\hat{m}}_{B-1}$,
and the channel output $\bml{y}^{(B-1)}$. It looks for an index $j$ such that
\begin{equation}
\label{eq:decoding_2}
(\bml{u}^{(B-1)}(\hat{\hat{m}}_{B-1}|j),\bml{v}^{(B-1)}(j),\bml{y}^{(B-1)}) \in {\cal T}_{UVY}
\end{equation}
Similarly, for $2\leq b\leq B-1$, the decoder looks for an index $j$ such that
\begin{equation}
\label{eq:decoding_3}
(\bml{u}^{(b)}(\hat{\hat{m}}_{b}|j),\bml{v}^{(b)}(j),\bml{y}^{(b)}) \in {\cal T}_{UVY}
\end{equation}
If such index $j$ exists, and is unique, the decoder sets $\hat{\hat{m}}_{b-1}=j$. Otherwise,
an error is declared.
Assuming that~$m_b$ was decoded correctly in the previous decoding stage. 
i.e., $\hat{\hat{m}}_b=m_b$, the decoding of $m_{b-1}$ in Sub-block $b$ is correct with probability close
to 1 provided that~(\ref{eq:decoding_rate_bound}) holds. Note that~$m_1$ is decoded in Sub-block $b=2$,
that is, $\bml{y}^{(1)}$ is not used at the destination. However, the transmission in Sub-block~1
is not superfluous, as it is used by the helper to decode
$m_1$ at the end of the first sub-block. Since~(\ref{eq:code_helper_end1}) 
and~(\ref{eq:decoding_rate_bound}) are the two terms in~(\ref{eq:def_ci}),
this concludes the proof of  the direct part.

\subsection{Converse Part}
Fix $|{\cal T}|$, and consider
$(n,2^{nR},|\set{T}|,\tilde{\epsilon}_n)$-codes with
$\tilde{\epsilon}_n \downarrow 0$. For each $n$, feed a randm message
$\Ma$ that is uniformly distributed on $\{1,2,\ldots,2^{nR}\}$ to the
encoder. By the channel model, 
\begin{equation}
\label{eq:markov_basic}
 \Ma \markov (X^nS^n)\markov Y^n.
\end{equation}
Fano's inequality and the fact that $\tilde{\eps}_{n} \downarrow 0$
imply the existence of a sequence $\eps_{n} \downarrow 0$ for which
\begin{IEEEeqnarray}{rCl}
n(R-\epsilon_n) &\leq& I(\Ma;Y^n) \stackrel{(a)}{\leq} I(\Ma;X^nS^n) = I(\Ma;X^n|S^n)\nonumber\\
  &=& \sum_{i=1}^n I(\Ma;X_i|S^nX^{i-1})\nonumber\\
  &\stackrel{(b)}{=}& \sum_{i=1}^n I(\Ma;X_i|S^nX^{i-1}T_i)\nonumber\\
  &\leq& \sum_{i=1}^n I(\Ma S_i^n;X_i|S^{i-1}X^{i-1}T_i)\nonumber\\
  &\stackrel{(c)}{=} &\sum_{i=1}^n I(\Ma;X_i|S^{i-1}X^{i-1}T_i)\nonumber\\
  &\leq& \sum_{i=1}^n I(\Ma Y^{i-1};X_i|S^{i-1}X^{i-1}T_i)\label{eq:rate_bound_helper1}
\end{IEEEeqnarray}
where $(a)$ follows from~\eqref{eq:markov_basic}; $(b)$ holds because $T_i$ 
is a function of $X^{i-1}S^i$~\eqref{eq:ti_definition}; and $(c)$
holds because $X_i$
is a function of $\Ma T^i$ and hence of $\Ma S^{i-1}X^{i-1}T_i$ (so
$I(S_{i}^{n};X_{i}|\Ma S^{i-1}X^{i-1}T_i)$ must be zero).

We proceed to derive the second bound. Starting again with Fano's inequality,
\begin{IEEEeqnarray}{rCl}
n(R-\epsilon_n) &\leq& I(\Ma;Y^n) =\sum_{u=1}^n I(\Ma;Y_i|Y^{i-1})\nonumber\\
 &\leq& \sum_{u=1}^n I(\Ma Y^{i-1};Y_i)\label{eq:rate_bound_decoder1}
\end{IEEEeqnarray}

Defining 
\begin{IEEEeqnarray}{rCl}
U_i&=& \Ma Y^{i-1}\label{eq:U_def}\\
V_i&=& S^{i-1}X^{i-1}\label{eq:V_def}
\end{IEEEeqnarray}
we can rewrite~\eqref{eq:rate_bound_helper1}
and~\eqref{eq:rate_bound_decoder1} as
\begin{subequations}
  \label{block:amos_block_ineq_n}
  \begin{IEEEeqnarray}{rCl}
R-\epsilon_n &\leq& \one\sum_{i=1}^n I(U_i;X_i|V_iT_i)\label{eq:rate_bound_helper2}\\
R-\epsilon_n &\leq& \one\sum_{i=1}^n I(U_i;Y_i)\label{eq:rate_bound_decoder2}
\end{IEEEeqnarray}
\end{subequations}
Moreover, with $U_{i}$ and $V_{i}$ defined as above, $U_{i}V_{i}$ and
$S_{i}$ are independent
\begin{equation}
  \label{eq:amos_indepUVS}
  \bigl( U_{i}V_{i} \bigr) \indep S_{i}
\end{equation}
and
\begin{IEEEeqnarray}{rCl}
T_i&=&h_i(S_i,V_i)\label{eq:function_ti}\\
X_i &=& f_i(U_i,V_i,T_i)\label{eq:function_xi}
\end{IEEEeqnarray}
where $h_i$ and $f_i$ are (blocklength-dependent) deterministic
functions. Indeed, $X_{i}$ can be determined from
$(U_{i},V_{i},T_{i})$ because $U_{i}$ determines the message $M$, and
$V_{i}$ determined $T^{i-1}$, so $(U_{i},V_{i},T_{i})$ determines
$(M,T^{i})$ from which $X_{i}$ can be computed using
\eqref{eq:def_encoder_intro2}.

We next do away with the sums by conditioning on a time-sharing
random variable: Let $Q$ be a random variable uniformly distributed over
$\{1,2,\ldots,n\}$, independently  of the channel and the state. Using
$Q$, we can express the bounds in~\eqref{block:amos_block_ineq_n} as
\begin{subequations}
  \label{block:amos_Rminus_inequalities}
\begin{IEEEeqnarray}{rCl}
R-\epsilon_n &\leq & I(U_Q;X_{Q}|V_QT_Q Q)\nonumber\\
                     &=&   I(U_Q Q;X_{Q}|V_QT_Q Q)\nonumber\\
                     &=& I(\tilde{U};X|VT) = I(\tilde{U}V;X|VT)\nonumber\\
                     &=& I(U;X|VT)\label{eq:rate_bound_helper3}\\
R-\epsilon_n &\leq& I(U_Q;Y_{Q}|Q)\nonumber\\
                     &\leq& I(U_Q Q;Y_{Q})\nonumber\\
                     &=& I(\tilde{U};Y)\leq I(\tilde{U}V;Y)\nonumber\\
                     &=& I(U;Y)\label{eq:rate_bound_decoder3}
\end{IEEEeqnarray}
\end{subequations}
where we define
\begin{equation}
  \label{eq:amos_XQ_def}
  X = X_{Q}, \qquad Y = Y_{Q}, \qquad T = T_{Q}, \qquad S = S_{Q}
\end{equation}
and the auxiliaries
\begin{IEEEeqnarray}{rCl}
V&=&(V_Q Q)\label{eq:ext_def_V}\\
 \tilde{U}&=&(U_Q Q)\label{eq:ext_def_tildeU}\\
 U&=&(\tilde{U},V) = (U_{Q}V_{Q}Q). \label{eq:ext_def_U}
\end{IEEEeqnarray}
Note that the conditional law of $Y$ given $(XTS)$ is that of the
channel, namely, $W_{Y|XS}$ and that $S$ is distributed like the channel state. Moreover,
\begin{equation}
V\markov U\markov (XTS)\markov Y.\label{eq:markov_UV}
\end{equation}

Since $U$ and $V$ contain the time sharing random variable $Q$,
\eqref{eq:function_ti} and~\eqref{eq:function_xi} imply that,
\begin{IEEEeqnarray}{rCl}
T &=& h(S,V)\label{eq:function_T}\\
X &=& \tilde{f}(\tilde{U},V,T) = f(U,T)\label{eq:function_X}
\end{IEEEeqnarray}
for some deterministic functions $h$ and $f$.  Therefore the joint
distribution under which the RHS of \eqref{eq:rate_bound_helper3} and
of \eqref{eq:rate_bound_decoder3} is of the form
\begin{equation}
P_{S\tilde{U}VTXY} = P_S \, P_{\tilde{U}V} \,
P_{T|SV}P_{X|\tilde{U}VT} \, W_{Y|XS}\label{eq:joint_dist_converse_tilde}
\end{equation}
where $P_{T|SV}$ and $P_{X|\tilde{U}VT}$ are $0-1$ laws, or
\begin{equation}
P_{SUVTXY} = P_S \, P_U \, P_{V|U} \, P_{T|SV} \, P_{X|UT} \, W_{Y|XS}\label{eq:joint_dist_converse}
\end{equation}
where $P_{T|SV}$, $P_{X|UT}$ and $P_{V|U}$ are $0-1$ laws.

The form~\eqref{eq:joint_dist_converse} and the inequalities in
\eqref{block:amos_Rminus_inequalities} establish the converse.

We next proceed to bound the alphabet sizes of $U,V$. In the first
step we do so by relaxing the $0-1$-law requirements. In the second
step will be further enlarge the alphabets to additionally satisfy
said requirements.

Let
\begin{equation}
  L= |{\cal X}| \, |{\cal T}| \ |{\cal S}| + 1.
\end{equation}
Fix a conditional distribution $p(x,t,s|u)$, and define the $L$
functions of $p(u|v)$:
\begin{IEEEeqnarray}{rCl}
p(x,t,s|v) &=& \sum_u p(x,t,s|u) \, p(u|v)\label{eq:constraints_1}\\
               & &\mbox{}   \quad\mbox{($L-2$ functions})\nonumber\\
& &I(U;X|T,V=v) \nonumber\\
& & I(U;Y|V=v)\nonumber
\end{IEEEeqnarray}
(with the $L-2$ functions corresponding to all by one of the tuples
$(x,t,s)$). By the support
lemma~\cite{CsiszarKorner:82b},\cite{ElGamalKim:11b}, there exists a
random variable $V'$ with alphabet $|{\cal V}'|\leq L$, such that
$P_{XTS}$, $I(U;X|TV)$ and $I(U;Y)$ are preserved. Denote by $U'$ the
resulting random variable $U$, i.e.,
\begin{equation}
P_{U'}(u') = \sum_{v'} p(u'|v)P_{V'}(v') \label{eq:Uprime_def}
\end{equation}
We next bound the alphabet size of $U'$. For each $v'\in{\cal V}'$ we define the $L$ functions
\begin{IEEEeqnarray}{rCl}
&& p(x,t,s|v',u')  \quad\mbox{($L-2$ functions})\label{eq:constraints_p_2}\\
& &I(U';X|T,V') \label{eq:constraints_iux_2}\\
& & I(U';Y|V')\label{eq:constraints_iuy_2}
\end{IEEEeqnarray}
Applying again the support lemma, for every $v'$ there exists a random variable $U''$ with alphabet $|{\cal U}''|\leq L$
such that~(\ref{eq:constraints_p_2}), (\ref{eq:constraints_iux_2}) and~(\ref{eq:constraints_iuy_2}) are preserved.
If we multiply ${\cal U}''$  $|{\cal V}'|$  times we can, with proper labeling of the elements of ${\cal U}''$, retain 
a Markov structure like~(\ref{eq:markov_UV}).
Now the alphabets sizes are fixed and independent of $n$. Thus, substituting $V',U''$ in~(\ref{eq:rate_bound_helper3}),
(\ref{eq:rate_bound_decoder3}) and taking the limit
$n\rightarrow\infty$ we have the upper bound
\begin{IEEEeqnarray}{rCl}
R &\leq& I(U'';X|V'T)\label{eq:rate_bound_helper4}\\
R &\leq& I(U'';Y) \label{eq:rate_bound_decoder4}
\end{IEEEeqnarray}
where
\begin{IEEEeqnarray}{rCl}
P_{SU''V'TXY} &=& P_S \, P_{U''V'} \, P_{T|SV'} \, P_{X|U''V'T} \, W_{Y|XS}\label{eq:joint_dist_converse2}\\
 & & |{\cal V}'|\leq L,\quad |{\cal U}''|\leq L^2 \label{eq:alphabet_sizes_Vp_Upp}
\end{IEEEeqnarray}
and the following Markov chain holds:
\begin{equation}
V'\markov U''\markov (XTS)\markov Y. \label{eq:markov_VpUpp}
\end{equation}

Note, however, that $P_{T|SV'}$ and $P_{X|U''V'T}$ are no longer $0-1$
laws. We remedy this using the Functional Representation lemma
(FRL)~\cite{ElGamalKim:11b}
(at the
cost of increasing the alphabets sizes): a standard convexity argument
will not do because---although $I(U;X|VT)$ is a convex function of
$P_{T|SV}$ and also a convex function of $P_{X|UVT}$ and likewise 
$I(U;Y)$---the minimum of two convex functions need not
be convex.

The Functional Representation lemma implies that---without altering
the conditional law of $T$ given $SV'$ nor of $X$ given
$U'' V' T$---the random variables $T$ and $X$ can be represented as
\begin{IEEEeqnarray}{rCl}
T &=& \tilde{g}_1(SV', Z_1)\label{eq:FRL_T_1}\\
X &=& \tilde{g}_2(U''V'T, Z_2)\label{eq:FRL_X_1}
\end{IEEEeqnarray}
where $\tilde{g}_1$, $\tilde{g}_2$ are deterministic functions; $Z_1$
and $Z_2$ are independent random variables that are independent
of $(SV', U''V'T)$; and their alphabets satisfy
\begin{IEEEeqnarray}{rCl}
|{\cal Z}_1| &\leq& |{\cal S}| \, |{\cal V}'| \, \left( |{\cal T}|-1
\right) + 1 \label{eq:FRL_Z1_size}\\
|{\cal Z}_2| &\leq& |{\cal U}''| \, |{\cal V}'| \, |{\cal T}| \, \left( |{\cal X}| - 1 \right) +1 \label{eq:FRL_Z2_size}
\end{IEEEeqnarray}
At the expense of increased alphabets sizes, we now append $Z_{1}$ to $V'$
and $Z_{2}$ to $U''$ to form the new auxiliary random variables
\begin{IEEEeqnarray}{rCl}
\hat{V} &=& (V'Z_1)\label{eq:hatV}\\
\hat{U} &=& (U''Z_2)\label{eq:hatU}
\end{IEEEeqnarray}
with alphabet sizes
\begin{IEEEeqnarray}{rCl}
|\hat{\cal V}| &\leq& |{\cal S}| |{\cal V}'|^2 \left( |{\cal T}|-1 \right) +|{\cal V}'|\label{eq:hatV_size}\\
|\hat{\cal U}| &\leq& |{\cal U}''|^2 |{\cal V}'| |{\cal T}| \left( |{\cal X}|-1 \right) +|{\cal U}''|\label{eq:hatV_size}
\end{IEEEeqnarray}
We set
\begin{equation}
  \label{eq:amos_map_to_x}
    P_{X|\hat{U} \hat{V} T}(x | u'', z_{2}, v', z_{1}, t)  = \indicatorfunction{\bigl\{x =
      \tilde{g}_{2}(u'', z_{2}, v', t)\bigr\}}
  \end{equation}
  (irrespective of $z_{1}$) and
  \begin{equation}
    P_{T|\hat{V}S}(t|v', z_{1}, t)  = \indicatorfunction{\bigl\{ t = g_{1}(s,v',z_{1}) \bigr\}}
  \end{equation}
  where $\indicatorfunction{\{\text{statement}\}}$ equals $1$ if the
  statement is true and equals $0$ otherwise.

  As we next argue, these auxiliary random variables and the above
  zero-one laws do not decrease the relevant mutual information
  expressions.

  Beginning with $I(\hat{U};X|\hat{V}T)$, we note that
  $H(X|\hat{V}T) = H(X|V'T)$ because we have preserved the joint law
  of $V'T$ and because $Z_{1}$ does not influence the
  mapping~\eqref{eq:amos_map_to_x} to $X$. Since
  $H(X|U''Z_{2}VT) \leq H(X|H(X|U''VT)$, this establishes that
\begin{equation}
  I(\hat{U};X|\hat{V}T) \geq I(U'';X|V'T).
\end{equation}

Likewise, our new auxiliary random variables and zero-one laws do not
alter $H(Y)$, but $H(Y|\hat{U}) \leq H(Y|U'')$,
so
\begin{equation}
  I(\hat{U};Y) \geq I(U'';Y).
\end{equation}

This completes the proof of Theorem~\ref{theo:main}.


\IEEEtriggeratref{5}

 \bibliographystyle{hieeetr}
 
  \bibliography{./references_Yossi.bib}

\end{document}